\documentclass{article}
\usepackage{spconf, amsmath, graphicx, bm, amssymb, multirow, lipsum, booktabs, subcaption, makecell, caption, url, enumitem, pifont}

\title{Retrieval Augmented Cross-Modal Tag Recommendation in Software Q\&A Sites}
\name{Sijin Lu$\mathbf{\dag}$\thanks{{\dag} Equal contribution.}, Pengyu Xu$\mathbf{\dag}$, Bing Liu, Hongjian Sun, Liping Jing*\thanks{* Corresponding author}, Jian Yu}
\address{Beijing Key Lab of Traffic Data Analysis and Mining, Beijing Jiaotong University, Beijing, China \\
\{22120406, pengyu, 22120391, 23120407, lpjing, jianyu\}@bjtu.edu.cn}
\begin{document}
%
\maketitle

\begin{abstract}
Posts in software Q\&A sites often consist of three main parts: title, description and code, which are interconnected and jointly describe the question. 
Existing tag recommendation methods often treat different modalities as a whole or inadequately consider the interaction between different modalities. Additionally, they focus on extracting information directly from the post itself, neglecting the information from external knowledge sources. Therefore, we propose a Retrieval Augmented Cross-Modal (RACM) Tag Recommendation Model in Software Q\&A Sites. 
Specifically, we first use the input post as a query and enhance the representation of different modalities by retrieving information from external knowledge sources. For the retrieval-augmented representations, we employ a cross-modal context-aware attention to leverage the main modality description for targeted feature extraction across the submodalities title and code.
In the fusion process, a gate mechanism is employed to achieve fine-grained feature selection, controlling the amount of information extracted from the submodalities. Finally, the fused information is used for tag recommendation.
Experimental results on three real-world datasets demonstrate that our model outperforms the state-of-the-art counterparts. 
\end{abstract}

\section{Introduction}
\label{sec:intro}
\begin{figure}[htb]
\centering
\includegraphics[width=8cm]{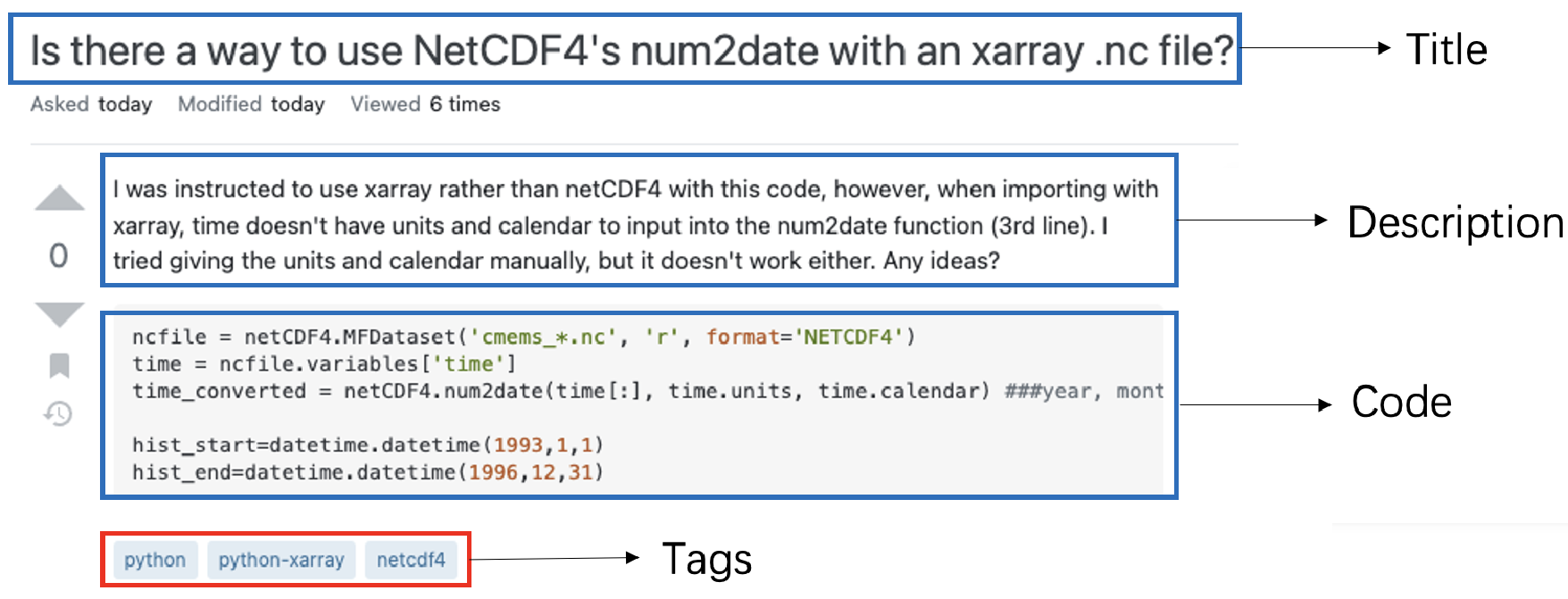}
\caption{An example in the Q\&A site StackOverflow.}\label{fig:post}
\end{figure}

Software Q\&A sites like StackOverflow have become essential resources for assisting software development \cite{wang2018entagrec++}. They primarily consist of various Q\&A posts, allowing users to tag their posts with one or more technical terms. A post typically comprises four components: title, description, code and tags, as illustrated in Figure ~\ref{fig:post}. 
The title can be considered a high-level summary of the post \cite{li2010semi}, serving as an overarching theme. The description further expands on the title, addressing the presented question and providing detailed explanations. While the code is closely linked to the description, providing essential programming information \cite{xu2021post2vec}. Therefore, the information from different modalities is closely interconnected, with the description being the most crucial part.
Previous works simply merged different modalities directly \cite{zhou2019deep, xu2016predicting} or even removed code during the preprocessing step \cite{wang2018entagrec++, treude2019predicting, yang2016security}, limiting the model's ability to fully capture diverse modality features. 
\cite{xu2021post2vec, he2022ptm4tag} independently learn the representations of each modality and integrate all the information for tag recommendation. \cite{li2023code} regards code as a semantic enhancement signal to enhance the representations of posts. 
However, the above-mentioned methods don't adequately taken into account the interactions between different modalities. Additionally, their fusion strategies are overly simplistic, neglecting the filtering of information.

Moreover, existing methods often rely solely on information extracted from the current post for prediction, disregarding relevant background or supplementary knowledge that can be obtained from external knowledge sources. Compared to parameterized training components, external knowledge sources are visible, easily modifiable and expandable without the need to retrain the model \cite{gur2021cross}, providing strong interpretability and scalability. Therefore, we construct a knowledge source of software Q\&A site with 4 million posts and employ retrieval augmentation to assist the tag recommendation task.

We propose a retrieval augmented cross-modal tag recommendation method, called RACM. The model utilize background and extended information to enhance the representations of posts by retrieving information from external knowledge sources. By leveraging the cross-modal context-aware attention and gate mechanism, the information from different modalities are effectively filtered and integrated, thereby enhancing the accuracy of tag recommendation.

The main contributions of this work are as follows:

\begin{itemize}[itemsep=0pt]
    \item[$\bullet$] We introduce the retrieval augmentation for tag recommendation for the first time and release a knowledge source consisting of 4 million software Q\&A posts.
    \item[$\bullet$] We employ the cross-modal context-aware attention and gate mechanism to achieve the interaction between modalities and perform information filtering.
    \item[$\bullet$] We provide three real-world datasets from software Q\&A sites to facilitate further research, each comprising title, description, code and tags.
    \item[$\bullet$] We perform experiments on three benchmark datasets. The F1@5 metric demonstrates an average improvement of 2.9\% over the best comparative algorithm.
\end{itemize}

\begin{figure*}[htb]
\centering
\includegraphics[width=0.8\textwidth]{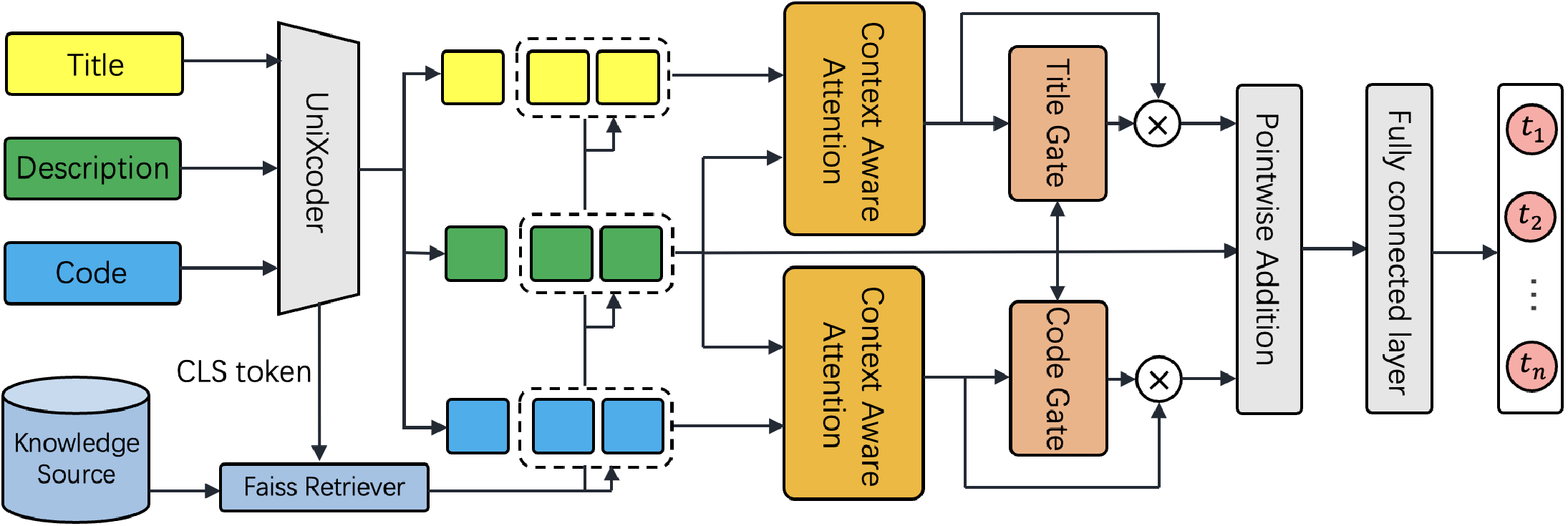}
\caption{Overview of RACM.}\label{fig:model}
\end{figure*}

\section{Method}
\label{sec:method}

As depicted in Figure~\ref{fig:model}, our RACM is composed of two major modules: retrieval augmentation and cross-modal interaction fusion. 
Specifically, retrieval augmentation enhances the representations of posts by retrieving information from the external knowledge source. 
Cross-modal interaction fusion involves the cross-modal context-aware attention and gate mechanism, enabling the interactive learning between modalities and the selective fusion of information.

\subsection{Problem Definition}
\label{ssec:subhead}

Let calligraphic letter (e.g., $\mathcal{A}$) indicate set, capital letter (e.g. $\mathrm{A}$) for scalar, lower-case bold letter (e.g. $\mathbf{a}$) for vector and capital bold letter (e.g. $\mathbf{A}$) for matrix. The input of the training stage includes $N$ instances $\mathcal{D}=\{(\mathbf{X}_i,\mathbf{y}_i)\}^N_{i=1}$, each of which consists of a post $\mathbf{X}$ and several tags $\mathbf{y} = (y_{1},...,y_{j},...,y_{l} )$ related to the post. Each post contains a title, description and code, $\mathbf{X} = \{(\mathbf{T}_i,\mathbf{B}_i,\mathbf{C}_i))\}^N_{i=1}$. Here $y_{j} \in \{ 0,1 \}$, where $y_{j} = 1$ indicates that the $j$-th tag is associated with post $\mathbf{X}$, and $l$ is the total number of candidate tags. In the testing stage, we aim to recommend the most relevant tags for a new post. 

\subsection{Initial Representation Learning}
\label{ssec:subhead}

We utilize the pre-trained model UniXcoder\cite{guo2022unixcoder}, which has demonstrated significant performance in code-related tasks as the encoder to learn the representations of all modalities:

\begin{equation}
[\mathbf{H}_t^s, \mathbf{H}_b^s, \mathbf{H}_c^s] = UniXcoder(\mathbf{X}).
\end{equation}

$\mathbf{H}_t^s, \mathbf{H}_b^s, \mathbf{H}_c^s \in \mathbb{R}^{n\times d}$ are the learned initial representations of the title, description and code, respectively. 
Here, $n$ denotes the maximum sequence length of the representations, and $d$ represents the feature dimension.

\subsection{Retrieval Augmentation}
\label{ssec:subhead}

We employ FAISS \cite{johnson2019billion} as our query index platform. We utilize the constructed external knowledge source as the target for retrieval, and our model has the capability to seamlessly switch to other knowledge sources.

Firstly, we obtain the CLS token representations $\mathbf{H}_{cls}^s$ of the input post learned by UniXcoder as query. Then, the retriever takes the query and the retrieval number $k$ as input and calculates the similarity using the Euclidean distance.
The knowledge source provides a collection of CLS token representations of posts $\{\mathbf{H}_{cls}^i\}^N_{i=1}$, where $N$ is the number of posts, the similarity calculation formula is as follows:

\begin{equation}
\mathrm{Sim}(\mathbf{H}_{cls}^s, \mathbf{H}_{cls}^i) = \sqrt{\sum_{j=1}^L({\mathbf{H}_{cls}^s}_j-{\mathbf{H}_{cls}^i}_j)^2}.
\end{equation}

Here, $L$ is the length of the post. The retriever returns $k$ title, description and code representations $[\mathbf{H}_p^1, \mathbf{H}_p^2,...,\mathbf{H}_p^k]$,
$p\in \{t,b,c\}$. Then we concatenate the representations of the retrieved posts with the initial representations:

\begin{equation}
\mathbf{H}_p = [\mathbf{H}_p^s \circ \mathbf{H}_p^1 \circ \dots \circ \mathbf{H}_p^k], p\in \{t,b,c\},
\end{equation}

\noindent where $\mathbf{H}_t, \mathbf{H}_b, \mathbf{H}_c \in \mathbb{R}^{(n*k)\times d}$ are the retrieval-augmented representations, which incorporate expanded information from the external knowledge source, 
improving the accuracy and robustness of the model.
Furthermore, during the training process, the retrieval process is dynamic and can be fine-tuned through indirect supervision from the loss, enabling the model to learn more accurate retrieval results.

\subsection{Cross-Modal Context-Aware Attention}
\label{ssec:subhead}

After obtaining the retrieval-augmented representations, we utilize the cross-modal context-aware attention \cite{yang2019context} for interactive learning between different modalities. As description contains the richest information, we treat description as the main modality and performs interactive learning with submodalities title and code. Firstly, query, key and value of description are generated:

\begin{equation}
[\mathbf{Q}, \mathbf{K}, \mathbf{V}] = \mathbf{H}_b[\mathbf{W}_Q, \mathbf{W}_K, \mathbf{W}_V],
\end{equation}

\noindent where $\mathbf{W}_Q, \mathbf{W}_K, \mathbf{W}_V \in \mathbb{R}^{d \times d}$ are learnable parameters to generate $\mathbf{Q}, \mathbf{K}, \mathbf{V} \in \mathbb{R}^{n \times d}$. 
To determine how much information to integrate from the code and how much to retain from the description, we introduce learnable parameters $\bm\lambda_k,\bm\lambda_v\in \mathbb{R}^{n \times 1}$ and $\mathbf{U}_k,\mathbf{U}_v\in \mathbb{R}^{d \times d}$. The learning process for key and value of code $\mathbf{K}_c$ and $\mathbf{V}_c$ is as follows:

\begin{equation}
\left[\!\!\!
\begin{array}{c}
    \substack{\displaystyle \mathbf{K}_c} \\
    \substack{\displaystyle \mathbf{V}_c}
\end{array}
\!\!\!\right] = (1 - 
\left[\!\!\!
\begin{array}{c}
    \substack{\displaystyle \bm{\lambda}_k} \\
    \substack{\displaystyle \bm{\lambda}_v}
\end{array}
\!\!\!\right]) \left[\!\!\!
\begin{array}{c}
    \substack{\displaystyle \mathbf{K}} \\
    \substack{\displaystyle \mathbf{V}}
\end{array}
\!\!\!\right] + \left[\!\!\!
\begin{array}{c}
    \substack{\displaystyle \bm{\lambda}_k} \\
    \substack{\displaystyle \bm{\lambda}_v}
\end{array}
\!\!\!\right](\mathbf{H}_c\left[\!\!\!
\begin{array}{c}
    \substack{\displaystyle \mathbf{U}_k} \\
    \substack{\displaystyle \mathbf{V}_v}
\end{array}
\!\!\!\right]).
\end{equation}

We use the sigmoid activation function to train the values of parameters $\bm{\lambda}_k$ and $\bm{\lambda}_v$. 

\begin{equation}
\left[\!\!\!
\begin{array}{c}
    \substack{\displaystyle \bm{\lambda}_k} \\
    \substack{\displaystyle \bm{\lambda}_v}
\end{array}
\!\!\!\right] = \sigma(
\left[\!\!\!
\begin{array}{c}
    \substack{\displaystyle \mathbf{K}} \\
    \substack{\displaystyle \mathbf{V}}
\end{array}
\!\!\!\right] \left[\!\!\!
\begin{array}{c}
    \substack{\displaystyle \mathbf{w}_{k_1}} \\
    \substack{\displaystyle \mathbf{w}_{v_1}}
\end{array}
\!\!\!\right] + \mathbf{H}_c\left[\!\!\!
\begin{array}{c}
    \substack{\displaystyle \mathbf{U}_k} \\
    \substack{\displaystyle \mathbf{V}_v}
\end{array}
\!\!\!\right] \left[\!\!\!
\begin{array}{c}
    \substack{\displaystyle \mathbf{w}_{k_2}} \\
    \substack{\displaystyle \mathbf{w}_{v_2}}
\end{array}
\!\!\!\right]),
\end{equation}

\noindent where $\mathbf{w}_{k_1}, \mathbf{w}_{v_1}, \mathbf{w}_{k_2}$, and $\mathbf{w}_{v_2} \in \mathbb{R}^{d \times 1}$ are learnable parameters. 
We use the same method to obtain the key and value of title $\mathbf{K}_t$ and $\mathbf{V}_t$. Then, we use the dot-product attention to generate new representations for title and code:

\begin{equation}
\mathbf{H}_{t}^{'}=\mathrm{Softmax}(\frac{\mathbf{Q}\mathbf{K}^T_t}{\sqrt{d_t}})\mathbf{V}_t,
\end{equation}

\begin{equation}
\mathbf{H}_{c}^{'}=\mathrm{Softmax}(\frac{\mathbf{Q}\mathbf{K}^T_c}{\sqrt{d_c}})\mathbf{V}_c.
\end{equation}

\subsection{Gate Mechanism}
\label{ssec:subhead}

We use a gate mechanism \cite{kumar2022did} to control the amount of information transmitted by each submodality. The title gate and code gate are as follows:

\begin{equation}
\mathbf{G}_{t}=[\mathbf{H}_{b}\oplus\mathbf{H}_{t}^{'}]\mathbf{W}_{t}+\mathbf{B}_{t},
\end{equation}

\begin{equation}
\mathbf{G}_{c}=[\mathbf{H}_{b}\oplus\mathbf{H}_{c}^{'}]\mathbf{W}_{c}+\mathbf{B}_{c},
\end{equation}

\noindent where $\mathbf{W}_{t}$ and $\mathbf{W}_{c}\in \mathbb{R}^{2d \times d}$ are weight parameters, and $\mathbf{B}_{t}$ and $\mathbf{B}_{c}\in \mathbb{R}^{n \times d}$ are bias parameters. The symbol $\oplus$ denotes the concatenation operation, and $\odot$ denotes the element-wise multiplication. The gate can finely filter information in submodalities, updating all elements within the matrix. We use the filtered title and code representations with the description representations to obtain the global representations $\mathbf{H}$:

\begin{equation}
\mathbf{H}=\mathbf{H}_{b}+\mathbf{G}_{t}\odot\mathbf{H}_{t}^{'}
+\mathbf{G}_{c}\odot\mathbf{H}_{c}^{'}.
\end{equation}

\subsection{Model Training and Inference}
\label{ssec:subhead}

After the fusion, the global representations $\mathbf{H}$ is mapped to the recommended tag vector through a fully connected layer.
Given a input post $\mathbf{X}$ and its corresponding tags $\mathbf{y}$,
we use corresponding binary cross entropy as the loss function for the model $f$:

\begin{equation}
\mathcal{L}=-\frac{1}{M}\sum^{M}_{i=1}\sum^{N}_{j=1}\mathbf{y}_{ij} \mathrm{log}(f(\mathbf{X}))+
(1-\mathbf{y}_{ij})\mathrm{log}(1-f(\mathbf{X})),
\end{equation}

\noindent where $M$ represents the number of posts in the training set, and $N$ represents the number of tags related to each post. Based on the probabilities generated by the model $f$, we recommend the top-k tags with the highest probabilities to users.

\section{Experiment}
\label{sec:experiment}

\subsection{Experimental Settings}
\label{ssec:subhead}

\begin{table}[t]
\begin{center}
\caption{\label{tab:dataset}Basic Dataset Statistics}
\begin{tabular}{c|cccc}
\hline
Datesets       &Posts  &Tags   &Avg.words  &Avg.Tags \\
\hline
SO  &60393  &10573  &240.55      &2.88     \\
AU  &50788  &2584   &141.61      &2.77     \\
CR  &18852  &880    &307.56      &2.62     \\
\hline
\end{tabular}
\end{center}
\end{table}

\textbf{Datasets.} We use three real-world datasets of different sizes: StackOverflow (SO), AskUbuntu (AU) and CodeReview (CR). SO dataset is released by \cite{he2022ptm4tag}. The XML files of AU and CR datasets are officially published and publicly available. We also released our processed datasets to facilitate related researches. The statistics of the three datasets are shown in Table~\ref{tab:dataset}. The number of posts in these datasets ranges from 18K to 60K, while the number of tags per post is around 3. 

\noindent\textbf{Metrics and Parameter Settings.} We use widely used metrics \cite{tang2019integral, gong2016hashtag}: Precision@k ($P@k$), Recall@k ($R@k$) and F1-score@k ($F1@k$) (k=1,5) to measure the performance. For all six metrics, the higher the better. For each dataset, the ratio of training, validation and testing sets is 8:1:1. In the experiments, the original learning rate is 1e-5 and batch size is 4. We determine the maximum sequence length based on the length of different modalities. We adopt the Adam optimizer and early stopping to stop the training process when the validation loss is no longer decreasing. The detailed parameter settings are released in our code.

\noindent\textbf{Baselines.} We compare RACM with six of the most representative baseline models: ABC \cite{gong2016hashtag}, ACN \cite{lei2020tag}, PBAM \cite{sun2018automatic}, Post2Vec \cite{xu2021post2vec}, PTM4Tag \cite{he2022ptm4tag}, and CETR \cite{li2023code}. ABC, ACN and PBAM treat all modalities as whole, incorporating attention mechanisms within neural networks to capture essential and distinctive features. 
Post2Vec and PTM4Tag independently learn different modalities and perform the fusion of all information.
CETR regards code as a enhancement signal to capture the semantic associations between tags and posts.

\label{ssec:subhead}

\begin{table}
\centering
\captionsetup{font=small} 
\renewcommand{\arraystretch}{0.8}

\caption{\label{tab:experiment}Performance Comparison on StackOverflow Dataset}
\begin{subtable}{0.5\textwidth}
\centering
\small
\resizebox{\textwidth}{!}{
\begin{tabular}{c|c|ccc|ccc}
\toprule
Dataset & Method & P@1 & R@1 & F1@1 & P@5 & R@5 & F1@5 \\
\midrule
\multirow{6}{*}{SO}
& ABC & 69.6 & 29.4 & 41.3 & 30.1 & 57.9 & 39.6 \\
& ACN & 65.0 & 27.3 & 38.5 & 28.0 & 53.5 & 36.7 \\
& PBAM & 62.0 & 25.7 & 36.4 & 26.8 & 51.5 & 35.2 \\
& Post2Vec & 71.3 & 30.8 & 43.0 & 28.1 & 54.7 & 37.1 \\
& PTM4Tag & 74.8 & 31.5 & 44.3 & 30.4 & 59.0 & 40.1 \\
& CETR & \underline{75.6} & \underline{31.9} & \underline{44.9} & \underline{32.6} & \underline{62.2} & \underline{42.8}\\
& RACM & \textbf{77.7} & \textbf{32.7} & \textbf{46.0} & \textbf{34.5}& \textbf{65.3} & \textbf{45.2} \\

\toprule
\multirow{6}{*}{AU}
& ABC & 56.6 & 23.2 & 32.9 & 28.2 & 53.0 & 36.8 \\
& ACN & 46.6 & 19.0 & 27.0 & 23.2 & 43.4 & 30.2 \\
& PBAM & 41.3 & 16.4 & 23.5 & 21.8 & 40.9 & 28.5 \\
& Post2Vec & 57.0 & 24.3 & 34.1 & 27.7 & 52.1 & 36.2 \\
& PTM4Tag & 64.4 & 26.6 & 37.6 & 30.8 & 57.9 & 40.2 \\
& CETR & \underline{66.5} & \underline{27.8} & \underline{39.2} & \underline{32.1} & \underline{60.3} & \underline{41.9} \\
& RACM & \textbf{69.0} & \textbf{28.6} & \textbf{40.5} & \textbf{34.6} & \textbf{64.4} & \textbf{45.0} \\

\toprule
\multirow{6}{*}{CR}
& ABC & 68.1 & 29.6 & 41.2 & 30.0 & 60.2 & 40.1 \\
& ACN & 59.2 & 24.8 & 35.0 & 23.1 & 45.1 & 30.5 \\
& PBAM & 51.2 & 22.2 & 31.0 & 21.6 & 43.4 & 28.8 \\
& Post2Vec & 71.2 & 31.8 & 44.0 & 29.0 & 56.7 & 38.4 \\
& PTM4Tag & 80.6 & 35.0 & 48.8 & 33.9 & 65.7 & 44.7 \\
& CETR & \underline{81.7} & \underline{35.8} & \underline{49.8} & \underline{36.6} & \underline{72.0} & \underline{48.5} \\
& RACM & \textbf{83.4} & \textbf{36.8} & \textbf{51.1} & \textbf{38.9} & \textbf{77.1} & \textbf{51.7}\\
\bottomrule
\end{tabular}
}
\end{subtable}
\end{table}

\subsection{Evaluation Results}
\label{ssec:subhead}

\noindent\textbf{Performance Comparison.} We show the main comparison results on three datasets in Table ~\ref{tab:experiment}. 
We observe that the proposed RACM generally outperforms the
compared methods on the three datasets. For example, on the F1@5
metric, the relative improvements of RACM are 2.4\%, 3.1\% and 3.2\% compared with its best competitors on` three datasets, respectively.
ABC, PBAM and ACN are all attention-based neural network models that leverage both local and global attention mechanisms for tag recommendation. But they treat all modalities as a unified entity, disregarding the unique characteristics of different modalities, which limits the performance of these methods.
Post2Vec and PTM4Tag employ CNN and pre-trained models respectively to learn features from different modalities, and then integrate these distinct modalities. CETR takes into account the significance of code and utilizes it to enhance the post representations. However, the aforementioned methods lack comprehensive consideration of the interactions between different modalities, and only focus on learning from the posts themselves.

While RACM enhances initial representations by retrieving external  knowledge sources, then employs the cross-modal context-aware attention and gate mechanism for effective interaction, information filtering and fusion.
Therefore, RACM achieves the best results, significantly outperforming all comparisons by a large margin.

\begin{figure}[htb]
\begin{minipage}[b]{.48\linewidth}
  \centering
  \centerline{\includegraphics[width=4.0cm]{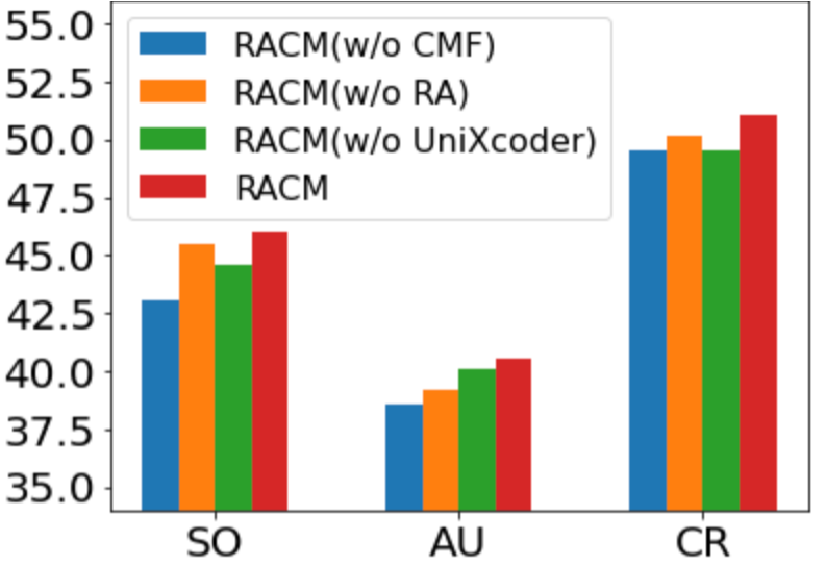}}
  \centerline{(a) F1@1 Result}\medskip
\end{minipage}
\hfill
\begin{minipage}[b]{0.48\linewidth}
  \centering
  \centerline{\includegraphics[width=4.0cm]{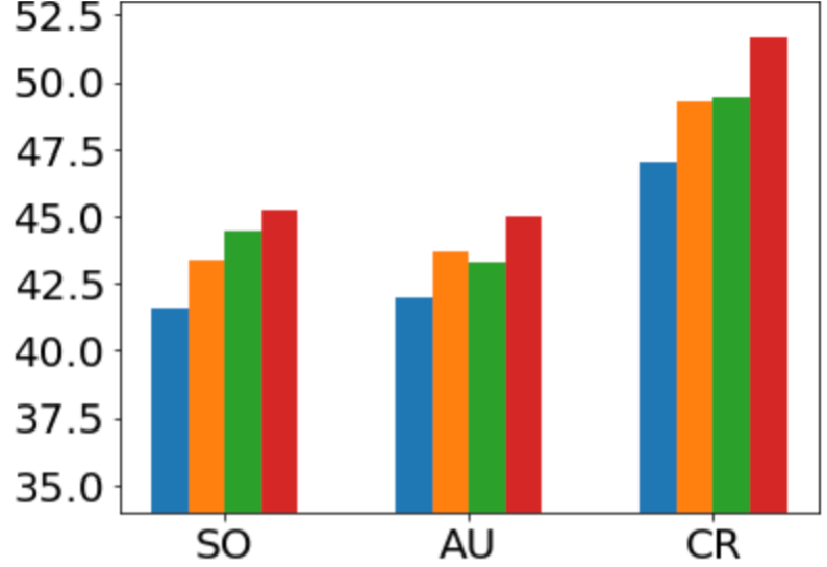}}
  \centerline{(b) F1@5 Result}\medskip
\end{minipage}
\caption{Results of Ablation Study.}
\end{figure}

\noindent\textbf{Ablation Study.} We analyze the impacts of key components in RACM via ablation study.
The default model is compared with the following variants.

RACM(w/o CMF) removes the cross-modal interaction fusion module. Compared with RACM(w/o CMF), RACM achieves higher 3.6\%, 3.0\% and 4.7\% F1@5 on three datasets respectively. It demonstrates the effectiveness of cross-modal context-aware attention and gate mechanism, which achieve the information interaction and selective fusion of different modalities, thus enabling better tag recommendation.

RACM(w/o RA) removes the retrieval augmentation module. Compared with RACM(w/o RA), RACM achieves higher 1.8\%, 1.3\% and 2.4\% F1@5 on three datasets respectively. It indicates that the retrieval augmentation is effective in enhancing post representations with background and supplementary knowledge.

RACM (w/o UniXcoder) replaces the cross-modal pre-trained model UniXcoder with BERT \cite{devlin2018bert}. F1@5 evaluation metric on three datasets decreased by 0.7\%, 1.7\% and 2.3\% respectively. It highlights the effectiveness of UniXcoder in handling code-related tasks.

\section{Conclusion}
\label{sec:con}

In this paper, we introduce RACM to address the tag recommendation problem. By retrieving information from the external knowledge source, more comprehensive and accurate post representations are obtained. The model employs cross-modal context-aware attention for interaction and utilizes gate mechanism to filter and fuse global information. Experimental results on three real datasets demonstrate that RACM significantly outperforms state-of-the-art models. 


\bibliographystyle{IEEEbib}
\bibliography{refs}

\end{document}